# Roughening Methods to Prevent Sample Impoverishment in the Particle PHD Filter

Tiancheng Li, Tariq P. Sattar
The Center for Automated and Robotics NDT
London South Bank University
London, SE1 0AA, UK
Lit3@lsbu.ac.uk; sattartp@lsbu.ac.uk

Qing Han, Shudong Sun
The School of Mechatronics
Northwestern Polytechnical University
Xi'an, 710072, P. R. China
hanqing008@163.com; sdsun@nwpu.edu.cn

*Abstract*— **Mahler's PHD (Probability Hypothesis Density) filter and its particle implementation (as called the particle PHD filter) have gained popularity to solve general MTT (Multi-target Tracking) problems. However, the resampling procedure used in the particle PHD filter can cause sample impoverishment. To rejuvenate the diversity of particles, two easy-to-implement roughening approaches are presented to enhance the particle PHD filter. One termed as "separate-roughening" is inspired by Gordon's roughening procedure that is applied on the resampled particles. Another termed as "direct-roughening" is implemented by increasing the simulation noise of the state propagation of particles. Four proposals are presented to customize the roughening approach. Simulations are presented showing that the roughening approach can benefit the particle PHD filter, especially when the sample size is small.**

## I. Introduction

The Probability Hypothesis Density (PHD) filter established by Mahler is a multiple target filter for recursively estimating the number and the state of a set of targets given a set of cluttered observations. It works by propagating in time the first moment of the multi-target posterior [1]. PHD filters implemented via weighted particles [2-3] (referred to particle filters, typically such as Vo's work [2], which is called the basic particle PHD filter in this paper) are relatively simple and free of linear and Gaussian requirement. Recently, some advanced particle implementations of the PHD filter have been proposed, such as stratified resampling [3] based on the weight component of particles, gating techniques [4] for fast computing and accurate estimation, and advanced technologies used in the particle filter that are extended into the particle PHD filter such as the auxiliary variable method [5], box-particle [6], and Rao-Blackwellisation implementation [7].

In general particle implementations of the PHD filter, the resampling procedure is a necessary and critical step that is not only for alleviating sample degeneracy as do in general particle filters but also to hard-limit the growth of the number of particles. It, however, can cause the notorious problem of sample impoverishment, leading to non-robust estimation in the particle PHD filter. In this paper, we investigate sample impoverishment caused by the resampling procedure and proposed two roughening strategies to rejuvenate the diversity of particles. Gordon's roughening procedure applied on particles after resampling [8] and a more direct roughening implementation that jitters the particle propagating dynamic are presented. The paper is organized as follows.

The basic framework of the particle PHD filter is reviewed in section II and based on it our roughening approaches are presented in section III. Simulation studies are given in section IV before we conclude in section V.

## II. Definitions and Problem Statement

### A. Mahler's PHD filter

To formulate the MTT filtering problem, the state of targets is generally assumed to follow a Markov process on the state space $\chi \subseteq \mathbb{R}^{nx}$, with transition density $f_{k|k-1}(\cdot|\cdot)$, i.e. given a state $x_{k-1}$ at time $k$-1, the probability density of a transition to the state $x_k$ at time $k$ is $f_{k|k-1}(x_k|x_{k-1})$. This Markov process is partially observed in the observation space $\mathbb{Z} \subseteq \mathbb{R}^{nz}$, as modelled by the likelihood function $g_k(\cdot|\cdot)$, i.e. given a state $x_k$ at time $k$, the probability density of receiving the observation $z_k \in \mathbb{Z}$ is $g_k(z_k|x_k)$. At time $k$, the collections of the states and measurements of targets can be represented as finite sets $X_k=\{x_{k,1}, \ldots, x_{k,Nk}\} \in F(\chi)$ and $Z_k=\{z_{k,1}, \ldots, z_{k,Mk}\} \in F(\mathbb{Z})$ respectively, where $N_k$ and $M_k$ are the number of targets and the number of measurements and $F(\chi)$ and $F(\mathbb{Z})$ are the collections of all finite subsets of targets and measurements, respectively.

Let $D_{k|k}$ and $D_{k|k-1}$ be the intensity functions associated to the posterior point processes: $x_k|Z_{1:k}=z_{1:k}$ and $x_k|Z_{1:k-1}=z_{1:k-1}$. We have the following Bayesian recursions [1, 2]

$$\ldots \to D_{k-1|k-1} \to D_{k|k-1} \to D_{k|k} \to \ldots$$

The transitions evolve via two operators: 1) prediction operator (time-update step)

$$D_{k|k-1} = \int_\chi \phi_{k|k-1}(x|u) D_{k-1|k-1}(u) du + \gamma_k(x) \qquad (1)$$

where the following abbreviation is used

$$\phi_{k|k-1}(x|u) = p_S(u) f_{k|k-1}(x|u) + b_k(x|u)$$

This work is supported by the National Natural Science Foundation of China (Grant no. 51075337). Tiancheng Li is also with the School of Mechatronics, Northwestern Polytechnical University, P. R. China.

where $b_k(x|u)$ denotes the intensity function of the RFS $B_k(x|u)$ of targets spawned from the previous state $u$, $p_S(x)$ is the survival probability of a target and $\gamma_k(x)$ is the birth intensity function of new targets at scan $k$.

2) updating operator (data-update step)

$$D_{k|k}(x) = (1-p_D(x))D_{k|k-1}(x) + \sum_{z \in Z_k} \frac{p_D(x)g_k(z|x)D_{k|k-1}(x)}{\kappa_k(z) + \int p_D(x)g_k(z|u)D_{k|k-1}(u)du} \quad (2)$$

where $g_k(z|x)$ is the single-sensor single target likelihood, $p_D(x)$ is the probability of detection and $\kappa_k(z)$ is the clutter intensity at time $k$. $\kappa_k(\cdot)$ can be written as $r_k c_k(\cdot)$, where $r_k$ is the average number of clutter points per scan and $c_k$ is the probability distribution of each clutter point.

Sets of target estimates are determined by obtaining peaks of the PHD. To address the computational issue, the particle method (commonly referred to as the particle filter) is proposed to approximate the PHD recursions [2]. Also, see more advanced implementation [3-7] and one stability study of the PHD filter [9]. In the following, we will present an enhanced particle PHD filter that aims to improve the diversity of particles after resampling.

*B. Particle implementation of PHD filters*

The particle filter propagates a set of particles with associated non-negative weights that approximates the probability density of the state according to Bayes' formula. It can be applied under very general hypotheses and is easy to implement. For the single-object case, the poster distribution of the state represented by particles can be written as

$$P(x_k) \approx \sum_{i=1}^{L_k} w_k^{(i)} \delta_{x_k^{(i)}}(x_k) \quad (3)$$

$$\sum_{i=1}^{L_k} w_k^{(i)} = 1, \quad w_k^{(i)} \geq 0 \quad (4)$$

where $\{x_k^{(i)}, w_k^{(i)}\}_{i=1,2,\dots,L_k}$ represent the states and weights of particles respectively, $L_k$ is the total number of particles at time $k$ and $\delta_x(\cdot)$ denotes the delta-Dirac mass located in $x$. The weights $w_k$ are chosen by using the principle of Sequential Importance Sampling (SIS), which relies on

$$w_k \propto w_{k-1} \frac{p(y_k|x_k)p(x_k|x_{k-1})}{q(x_k|x_{1:k-1}, y_k)} \quad (5)$$

where $q(\cdot)$ is a proposal importance density.

The particle PHD filter [2] uses the particles to approximate the PHD predictor and corrector that represent the first moment of the multi-target prior and posterior. The sum of the weights is no longer one but instead it approximates the expected number of targets, $\tilde{N}_k$

$$\tilde{N}_k \cong \sum_{i=1}^{L_k} w_k^{(i)}$$

To extract estimates, additional approximation (e.g. rounding operation) is required to get the integral number of targets.

Given the importance densities $p_k(\cdot|Z_k)$, $q_k(\cdot|x_{k-1}, Z_k)$ and supposing that there are $L_{k-1}$ particles in time step $k$-1 and $J_k$ new particles are allocated for possible new-born targets, according to (1) and (2), the particle implementation of the PHD predictor can be written as

$$D_{k|k-1}(x_k) = \sum_{i=1}^{L_{k-1}+J_k} w_{k|k-1}^{(i)} \delta_{x_k^{(i)}}(x_k) \quad (6)$$

where

$$x_k^{(i)} \sim \begin{cases} q_k(\cdot|x_{k-1}^{(i)}, Z_k), i=1,\dots,L_{k-1} \\ p_k(\cdot|Z_k), i=L_{k-1}+1,\dots,L_{k-1}+J_k \end{cases} \quad (7)$$

$$w_{k|k-1}^{(i)} = \begin{cases} \dfrac{\phi_{k|k-1}(x_k^{(i)}|x_{k-1}^{(i)})w_{k-1}^{(i)}}{q_k(x_k^{(i)}|x_{k-1}^{(i)}, Z_k)}, i=1,\dots,L_{k-1} \\ \dfrac{\gamma_k(x_k^{(i)})}{J_k p_k(x_k^{(i)}|Z_k)}, i=L_{k-1}+1,\dots,L_{k-1}+J_k \end{cases} \quad (8)$$

The particle implementation of the PHD corrector can be written as

$$D_{k|k}(x_k) = \sum_{i=1}^{L_{k-1}+J_k} w_{k|k}^{(i)} \delta_{x_k^{(i)}}(x_k) \quad (9)$$

where

$$w_{k|k}^{(i)} = \left[1 - p_D(x_k^{(i)}) + \sum_{z \in Z_k} \frac{p_D(x_k^{(i)})g_k(z|x_k^{(i)})}{\kappa_k(z) + C_k(z)}\right] w_{k|k-1}^{(i)} \quad (10)$$

with

$$C_k(z) = \sum_{j=1}^{L_{k-1}+J_k} p_D(x_k^{(j)}) g_k(z|x_k^{(j)}) w_{k|k-1}^{(j)} \quad (11)$$

The variance of the resulting estimates in particle filters will increase during the propagating of particles that will cause the known sample degeneracy. To mitigate this, resampling is generally applied which also provides an opportunity to adjust the number of particles. This is of significance to limit the growth of the number of particles in the particle PHD filter. However, after resampling most particles will have very similar or even the same states, leading to a further problem namely sample impoverishment. Particle degeneracy and impoverishment are similar problems appearing as unbalance between the need for diversity and the need for concentrate [10] that is arguably one of the fundamental difficulties of particle filters. In particular, sample impoverishment may occur in the particle implementation of the PHD filter as well and it requires the same level of attention.

Relevantly, the so-called weight over-estimate problem is pointed out in [11], which suggests that when the variance of measurement noise is small, sample degeneracy may become serious. This will further undermine the diversity of particles since most particles will be abandoned during the resampling. In this case, it becomes more necessary to rejuvenate the diversity of particles. Precisely because of these, we focus on the particle diversity in the particle PHD filter. Some solutions reported in the community are reviewed in the next section before our solutions are given.

## III. ROUGHENING PARTICLE PHD FILTER

### A. The state of the art

There are some strategies proposed by the particle filter community to combat the sample impoverishment caused by the resampling method. One idea is selective resampling which only resamples when necessary by monitoring the variance of important weights instead of always resampling. It can combat sample degeneracy while preventing sample impoverishment. However, the variance of weights is not only a manifestation of degeneracy as in the single object case. Since the multi-target distribution is arbitrary and unknown, the weight of particles distributed in different regions can be greatly different that cannot be attributed to degeneracy. For example, in the region of two or more target crossing (or being close), the weight of particles is naturally much higher than that of particles in the region with few or no target. More importantly, resampling is required iteratively to limit the growth of the number of particles, since in each iteration new particles are introduced to represent the new-born targets. Therefore, applying the selective resampling strategy based on the variance of the particle weights is actually not so straightforward for the particle PHD filter.

Drawing on ideas from the auxiliary particle filter, an auxiliary particle implementation of PHD filter is presented in [5] to reduce the variance of the importance weights of the particle PHD filter. The employed auxiliary method preselects particles for propagation on the basis of how well-matched they are to the next observation. The resulting algorithm samples are in a higher dimensional space than the basic particle PHD filter, which may lead to inefficiency in real-time performance not to mention the additional computation of the auxiliary variable. Similarly, other resampling methods e.g. deterministic resampling [10] have been proposed to avoid sample impoverishment with additional computation that can be huge. In the following section, we introduce two easy-to-implement and quite efficient strategies.

Since the so-called roughening strategy was first proposed by Gordon in the Bootstrap Filter [8] (called as dithering in [12]), one general idea to rejuvenate the diversity of particles after resampling is to increase the state noise covariance or to introduce an additional noise to the samples. Gordon's roughening strategy basically adds to each resampled particle an independent Gaussian jitter noise. The jitter noise, say $r_k$, is normally with zero mean and constant covariance $P_r$. The standard deviation is suggested in [8] as $KEN^{1/d}$, where $E$ is the difference between the maximum and minimum values of the state component, $K$ is a positive tuning constant chosen subjectively by the user, $N$ is the number of particles and $d$ is the dimension of the state. This is straightforward to incorporate in the particle PHD filter. Further, it is more desirable to use a dynamic tuning parameter $K$ for a certain case such as in reentry vehicle tracking [13].

### B. Proposed roughening methods

Suppose the original intensity represented by the resampled particles is denoted as $D_{k|k}$ (whose integral on any region $S$ of state space is the expected number of targets contained in $S$) [1]. Since the addition of two independent random variables corresponds to a convolution operation in the density domain, the intensity obtained after the roughening process can be factored as

$$\hat{D}_{k|k} = \langle D_{k|k}, r_k \rangle \quad (12)$$

where $\langle \cdot, \cdot \rangle$ denotes the convolution operation.

In fact, the roughening strategy could be implemented more directly by increasing the simulation noise of the dynamic propagation of particles. Suppose the system dynamic is $x_k = f_{k|k-1}(x_{k-1}, v_{k-1})$ where $v_{k-1}$ denotes the stochastic noise affecting the system dynamic equation $f_{k|k-1}$, the particle propagating dynamic perturbed with a zero-mean jitter can be written as

$$x_k^{(i)} = f_{k|k-1}(x_{k-1}^{(i)}, v_{k-1} + r_k) \quad (13)$$

It can be easily seen that the direct-roughening approach needs no additional online computation, while the online separate-roughening procedure is no more than a random variable generation. That is to say, both of them are quite computationally cheap and easy to implement. They have the equivalent effect of spreading the particles wider after resampling only differing at that the total spreading noise $v_{k-1}+r_k$, is executed separately in two times in the former roughening while jointly in one time in the latter. As a result, the sum of two independent normally distributed random variables is normal, with its mean being the sum of the two means, and its variance being the sum of the two variances.

The separate-roughening approach may enjoy smoother particle distribution at the price of (quite little) additional online computation as compared with the direct-roughening. In this paper to combat the sample impoverishment in the basic particle implementations, we firstly implement the separate-roughening procedure after resampling. Secondly, we adopt a bigger variance for the particle propagating dynamic. They are termed as separate-roughening approach and direct-roughening approach respectively. To note, there are several aspects involved with the implementation of roughening that could be improved

**Proposal 1**. Roughening may not be applied in all but only some recursions. This can be based on the variance monitoring of the variance of particles like selective resampling.

**Proposal 2**. Roughening may not be applied to all particles but only on 'overlapped' particles that are resampled from the same particle.

**Proposal 3**. Roughening may not be applied in all dimensionalities but only in a part of it, e.g. in the target tracking case, roughening is implemented only in the velocity space in our simulation.

The roughening approach is in fact making use of kernel smoothing. As with density estimation there needs to be a compromise between over and under smoothing the data. Adding noise or increasing the state noise variance only works under the premise that the state dynamic noise is relatively too small. More importantly, the sample impoverishment is arguably customized to specific situations as it may not always be serious or even not occur, as shown later in the simulations. Overall, the sample impoverishment is still a difficulty in the single target application of particle filters. For the roughening strategy in target tracking case, we suggest a proposal as follows to determine the

**Proposal 4**: Projected onto the target position space, the roughening magnitude corresponding to the roughening variance $\Sigma r_k$ is suggested to be no bigger than that of the (minimum, if there is more than one sensor) measurement noise variance.

Proposal 4 places an upper threshold on the roughening noise $r_k$. It causes the measurement noise to scale the roughening strength so that it would not blur the position estimates out of the range of the efficient observation. If the measurement is not directly made on the position of targets, e.g. bearing sensors receiving bearing measurements, it may need to be projected/mapped onto the target position space to determine the variance used for zero-mean roughening noises.

## IV. SIMULATIONS

Without loss of generality, we use the same simulation model as in [2]. Each target moves according to the following linear Gaussian dynamics

$$x_k = \begin{bmatrix} 1 & T & 0 & 0 \\ 0 & 1 & 0 & 0 \\ 0 & 0 & 1 & T \\ 0 & 0 & 0 & 1 \end{bmatrix} x_{k-1} + \begin{bmatrix} T^2/2 & 0 \\ T & 0 \\ 0 & T^2/2 \\ 0 & T \end{bmatrix} \begin{bmatrix} v_{1,k} \\ v_{2,k} \end{bmatrix} \quad (14)$$

where $x_k=[x_{1,k}, x_{2,k}, x_{3,k}, x_{4,k}]^T$, $T=1$ is the sampling time. The process noise $\{v_{1,k}\}$, $\{v_{2,k}\}$ are mutually independent zero-mean Gaussian white noise with respective standard deviation $\delta_{v1}=1$ and $\delta_{v2}=0.1$. Targets can appear or disappear in the scene at any time. The birth intensity of new targets is defined as $\gamma_k=0.2N(.; \bar{x}, Q)$, where $\bar{x} = [0, 3, 0, -3]^T$, Q= $diag([10, 1, 10, 1]^T)$, where $diag(a)$ gives a diagonal matrix in which the diagonal is $a$.

The target-originated range measurements are given by

$$y_k = \begin{bmatrix} 1 & 0 & 0 & 0 \\ 0 & 0 & 1 & 0 \end{bmatrix} x_k + \begin{bmatrix} w_{1,k} \\ w_{2,k} \end{bmatrix} \quad (15)$$

with $\{w_{1,k}\}$ and $\{w_{2,k}\}$ mutually independent zero-mean Gaussian white noise with standard deviations $\delta_{w1} = \delta_{w2} =2.5$. Clutter is uniformly distributed over the region $[-100,100]\times[-100,100]$ with an average rate of $r$ points per scan, i.e. $\kappa = r/200^2$. We use $r=10$. Each existing target has a (state independent) probability of survival $P_S(x) = 0.95$ and a probability of detection $P_D(x)= 0.95$.

To compare different filters, we adopt the OSPA metric [14]. For finite nonempty subsets $X=\{x_1, …,x_m\}$ and $Y=\{y_1, …, y_n\}$ of a closed and bounded observation window in $R^N$, the OSPA distance between $X$ and $Y$ is defined as (if $m\leq n$)

$$\bar{d}_p^{(c)}(X,Y) = \left( \frac{1}{n} \left( \min_{\pi \in \Pi_n} \sum_{i=1}^{m} d^{(c)}(x_i, y_{\pi(i)})^p + c^p(n-m) \right) \right)^{1/p} \quad (16)$$

If $m>n$, then $\bar{d}_p^{(c)}(X,Y) = \bar{d}_p^{(c)}(Y,X)$. In our case, the cut-off parameter $c=100$, the OSPA metric order parameter $p=2$.

To capture the average performance, we run 100 trials in all simulations with different target tracks and independently generated measurements. The tracking scene in the *x-y* plane is given in Fig.1 which shows the true trajectories of four targets and estimates of different filters ($N_p=1000$). The trajectories of targets and observations against the steps are plotted in *x* and *y* dimension respectively in Fig.2. For the simplicity, the importance sampling density used in (8) are the systematic dynamics $q_k=f_{k|k-1}$. The number of targets is estimated by the rounding calculation on the particle weight sum and the Bayesian multi-estimate extraction method [15] is used to determine state estimates of targets. $N_p$ particles per expected target are used in the simulation and the total number of particles is hard-limited so that it does not fall below $N_p/2$ even when the expected number of target is less than 0.5.

### A. Different degrees of sample impoverishment

To spread the resampled particles, a one-dimensional roughening noise is introduced in the velocity space only (see Proposal 3) which will be integrated to the position in the next iteration. We first choose the standard deviation of the zero-mean roughening noise $\delta_r=0.4$ which satisfies Proposal 4 by $\delta_r<\min(\delta_{w1}, \delta_{w2})$. As stated, the jitter noise is individually added into resampled particles in the separate-roughening approach while differently in the direct-roughening approach, it is incorporated into the propagation noise of particles. Both can be applied on part of particles according to Proposal 2.

Different numbers of particles are used in the simulation. The average estimation of the number of targets and average OSPA of different filters when $N_p=1000$ and $N_p=200$ are given in Fig. 3 and Fig. 4 respectively. For comparison, we define the gain ratio as the reduction percentage of the average OSPA by the roughening approach as compared with the basic particle PHD filter, which is calculated by

$$r_{\overline{OSPA}} = \frac{\overline{OSPA}_{basic} - \overline{OSPA}_{roughening}}{\overline{OSPA}_{basic}} \quad (17)$$

where $\overline{OSPA}$ denotes the mean of the average OSPA distances over 40 simulation steps of 100 trials.

As shown in the results, when $N_p=1000$ in the roughening approaches, it comes as no surprise, that no obvious advantage is obtained. The gain ratio is 6.2% and 5.5% on average respectively in the separate- roughening and direct-roughening improved particle PHD filters. When the sample size $N_p=200$, the influence of sample impoverishment becomes significant and the roughening filters has obtained a more accurate estimation of the number of target and a smaller miss-distance between the estimates and the true target states. On average, the gain ratio is 17.07% and 17.14% respectively in the separate roughening and direct-roughening improved particle filters. We conjecture this is because the sample impoverishment is not obvious or does not have obvious impact in the case of big sample size ($N_p=1000$) but is obvious when the sample size is relatively small.

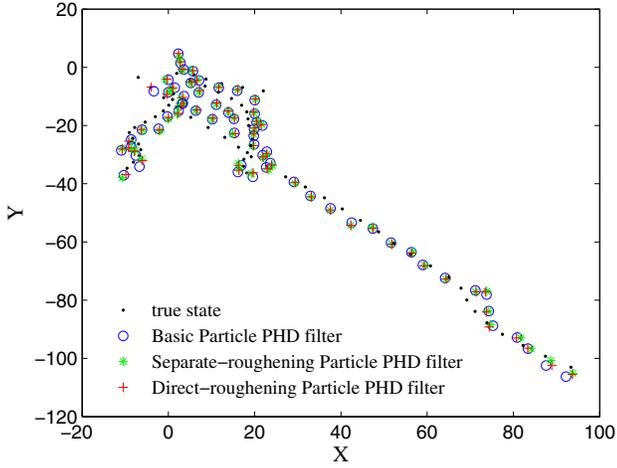

Fig.1    Tracking scene in one trial

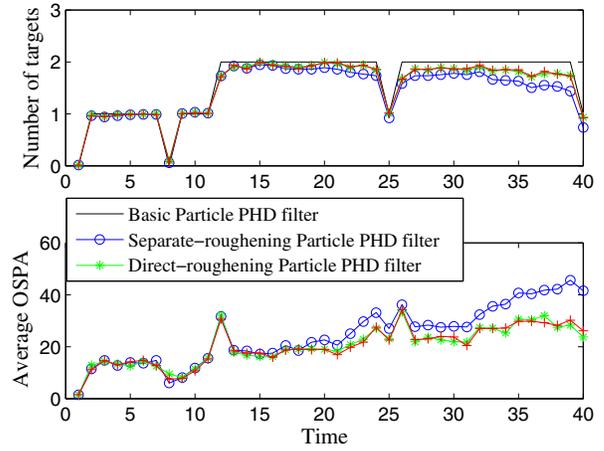

Fig. 4    Average estimated number of targets and OSPA when $N_p$=200

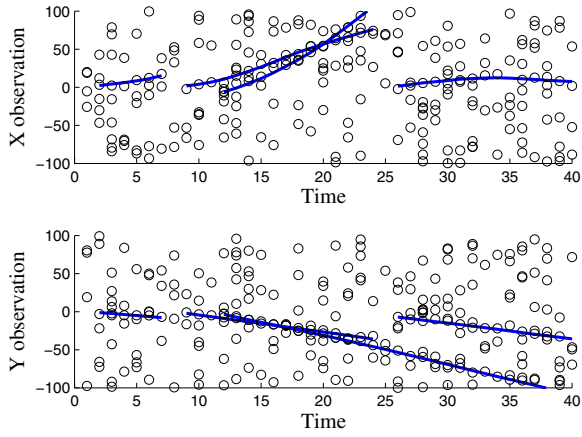

Fig.2    Trajectories of targets (blue line) and observations (black 'o')

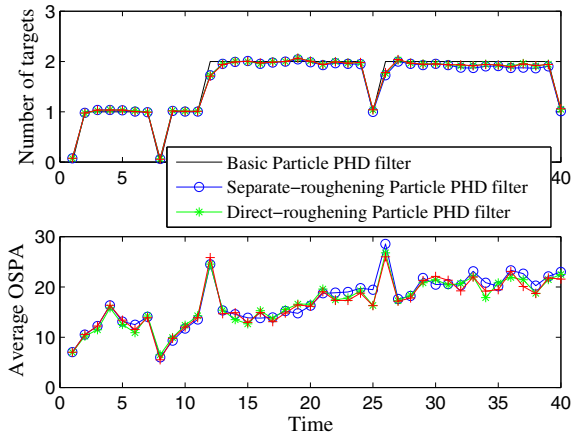

Fig. 3    Average estimated number of targets and OSPA when $N_p$=1000

## B. Different degrees of rougening

In this section, we apply different degrees of roughening noise to find relatively the best choice. $N_p$=200 particles per expected target are used. The average gain ratios against the standard deviation $\delta_r$ of the zero-mean roughening noise are plotted in Figure 5. The results indicate that the estimation is most improved by roughening strategies when $\delta_r$=[0, 0.2]. When the roughening noise is bigger or smaller than that, estimation is less improved and even reduced.

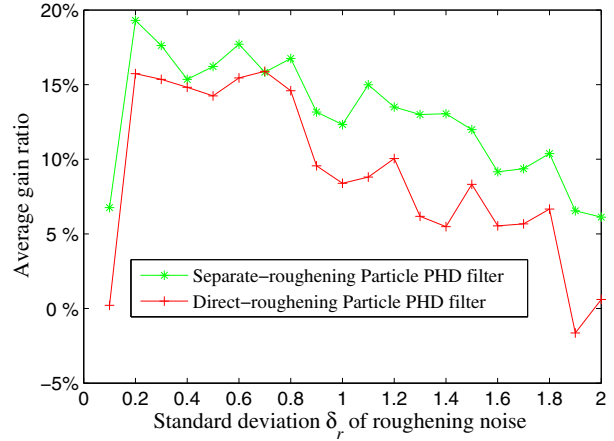

Fig. 5    The average gain ratio of roughening approaches

The simulations indicate that the solution to sample impoverishment is highly problem specific. First, sample impoverishment may not appear or is not obvious when a large number of particles are used. Also, advanced resampling techniques may be applied preventing sample impoverishment. In both cases, roughening is not necessary. Further, in the case of sample impoverishment, it is highly necessary to customize the degree of roughening to that of the impoverishment. Second, as stated over roughening (too big a roughening noise) usually leads to very dispersive particles and the additional noise will reduce the estimation accuracy while too small roughening is useless. In our approach, an offline searching through pre-simulation of the optimal roughening degree (including the zero variance, which means that it is better not to do roughening) is highly recommended within the scope limited by our Proposal 4.

V. CONCLUSION

As a critical step in the particle implementation of the PHD filter, resampling has major theoretical benefits (e.g. solving sample degeneracy) and practical benefits (e.g. limiting the growth of the number of particles), but it can cause sample impoverishment as well. Two easy-to-implement roughening strategies are proposed in this paper to rejuvenate the diversity of particles and to combat the possible sample impoverishment in particle PHD filters. Four proposals are presented to direct the roughening.

To implement the roughening strategy, it is necessary to customize to specific cases. In the case of obvious sample impoverishment e.g. when the sample size is relatively small, the proposed roughening strategies can obviously improve the performance of the particle PHD filter. The proposed roughening strategies are straightforward to enhance other particle implementations of PHD filters.

The simulation results of the direct roughening approach expose that the Sampling Importance Resampling (SIR) particle filter tends to have an inherent bias in estimation of the dynamic noise. As shown, a bigger dynamic noise (than the systemic one) used for the particle preparation will yield better filtering accuracy. Therefore, if the state dynamic noise needs to be estimated by using the SIR PF (namely parameter estimation), the estimate of the dynamic noise (obtained from particles) will tend to be bigger than the truth. This study will be a part of our future work.

ACKNOWLEDGMENT

Tiancheng Li would like to thank China Scholarship Council for supporting his research in United Kingdom.